# Security Survey of Internet Browsers Data Managers

## Catalin Boja

*Computer Science Department*
*Academy of Economic Studies*
*Romana Square no.6, Bucharest*
*ROMANIA*
*catalin.boja@ie.ase.ro*
*http://www.ism.ase.ro*

**Abstract:** The paper analyses current versions of top three used Internet browsers and compare their security levels to a research done in 2006. The security is measured by analyzing how user data is stored. Data recorded during different browsing sessions and by different password management functions it is considered sensitive data. The paper describes how the browser protects the sensitive data and how an attacker or a forensic analyst can access it.

## 1. Introduction

Accordingly with browser market share, [1], the top three used browsers are Microsoft Internet Explorer, Mozilla Firefox and Google Chrome.

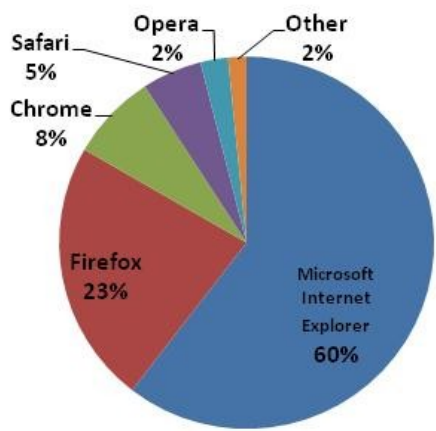

*Figure 1. Browser market share, Q3 2010*

The paper analyses current versions of top three Internet browsers and compare their security levels to the one described in a 2006 paper, [2] *Password Management Concerns with IE and Firefox* that analyzed Microsoft IE 6 and 7 and Firefox p1.5 and 2.0. In that research, the author defined and analyzed password storage mechanisms, attacks on password managers, the false sense of security, usability and the mitigation and countermeasures.

Most of the time, the entire security framework of the personal computer is based on the user management routines implemented by the operating system. Many home users base their computer security on user accounts to protect against physical or remote access to the machine. This is a dangerous and deprecated opinion because today IT security threats use the Internet and the user is more vulnerable when surfing or using Web services than not being at the computer. Users access more and more Web services, like social networking, Internet banking, online shopping and gaming that require sensitive personal data. And, the Web gate to all these resources is represented by the Web browser used to access the pages.

Today browsers have become complex software applications. It allows users to do more than just viewing web pages, adding up to their usability level through many other functions. One of these functions is to manage user passwords for different sites. The need for this option come from the fact that Web developers implement in their applications secure sessions that require users to authenticate with their username and password. And this is a reality for almost any type of Website, from a forum to an e-commerce one. So users must manage a set of multiple credentials used to authenticate on different websites.

As almost all users won't memorize multiple long and hard-to-guess passwords or write them on paper, the solutions that they may adopt are:
▪ use same password for multiple sites or applications;
▪ use password managers to store sensitive data.

The first solution is to be avoided because a single piece of information, represented by the used password, becomes the most vulnerable point of the personal data system, [2]. Acquiring that, an attacker will have access to all the Websites, e-mail, FTP accounts and other online services.

In Internet there are four primary ways users lose control over their web-based passwords, [15]:
▪ Phishing Scams – through fake websites or emails the attacker ask users to send him sensitive data like bank accounts, card PINs, passwords;
▪ Malware – users are convinced by the attacker to install malicious software, to access bad links and to allow malware to install on its machine through its browser; the malware can be a simple java script like a booklet password [20] or a more advanced third party software.

```
javascript:(function() {
var s,j,form,i; pass = "";
for(j=0; j<document.forms.length; ++j) {
form = document.forms[j];
     for (i=0; i<form.length; ++i) {
if (form[i].type.toLowerCase() == "password")
          pass += form[i].value + "\n";}}
if (pass) alert("Passwords on this page:\n\n" + pass);
else alert("There are no passwords in forms on this page.");
})();
```
*Figure 2. Booklet that shows the content of password fields*

▪ Website break-ins – the attacker succeeds to break the website security by SQL Injection, Remote File Inclusion (RFI), Cross Site Scripting (XSS), [13-14] or any other security vulnerability or misconfiguration and he manages to gain sensitive data or even to modify pages content in order to control users;
▪ Website brute-force attacks – the attacker breaks users' authentication credentials by trying possible passwords as in dictionary attacks.

The first two attacks and the XSS are strong related to browsers because they are based on scenarios in which the user uses the browser to open a fake webpage or to install malicious code. Using the browser is created a direct connection between users and attackers.
Because, it is possible for someone to gain full control from the Internet to the target machine, all possible attacks that can be conducted directly on the machine can also be used by the remote attacker.

## 2. Browsers sensitive data

The tight competition in the browsers market has motivated developers to offer more than a tool for just viewing Web pages. The quality of this type of software application can be analyzed from the viewpoint of these ISO 9126, [24], characteristics:
▪ efficiency measured as virtual memory needed to run the application and as speed in opening full downloaded Web pages;
▪ usability that describes how easy is to understand it and use it;
▪ functionality that defines the set of implemented functions;
The last two characteristics are materialized by including different tools and functions that help users to do with ease tasks as completing Web forms, creating secure connections, managing Web pages, debugging. All these advantages reduce the user effort to remember passwords, Web pages addresses and to complete Web forms with personal data, but also generate security vulnerabilities.

Data saved and managed by browsers is considered sensitive because contains:

▪ Autocomplete form data that represents phone numbers, addresses, user names, emails;

▪ Browsing history collecting previous visited URLs that describes a personal experience regarding what are the users informational needs;

▪ Bookmarks representing favorites links;

▪ Autocomplete passwords that give access to restricted Web resources and services;

▪ Passwords for FTP sites

## 3. Firefox security

From the 3.5 version, Mozilla Firefox, stores passwords in two different files, [3-5],

▪ key3.db, a binary file with a database table format, that stores the master key, the master password [28], for the passwords database; information in this files describes encrypted string, salt, algorithm and version information;

▪ signons.sqlite, a binary file with a SQLite format, [3], that stores the password and the username saved for different sites; the .sqlite file represents a complete SQL database being a cross-platform portable solution and this allow open access to the database structure; the database contains two tables, moz_disabledHosts used to store the URL of websites for which users have disable the password save function and moz_logins used to store passwords;

Figure 3 describes the process of encrypting and decrypting the passwords by Firefox in 3 steps:

1. The user access the Web page that uses a HTML authentication form;

2. After authentication, if the user chooses to save the password and username, Firefox stores the page URL, in clear text, in signons.sqlite file; the password and the username are encrypted with the master password from key3.db and Base64 encoded; if the user does not provide a master password, Firefox uses a default value for it, [4]; for encryption, Firefox uses 3DES in CBC mode, implemented by the PKCS #11, cryptographic API defined by RSA [25] and the cryptographic library Mozilla NSS [26]; if necessary, the user can configure Firefox to use a more secure encryption method defined by the Federal Information Processing Standard (FIPS) 140-2, [27];

3. The encrypted form of the password and username are stored in signons.sqlite.

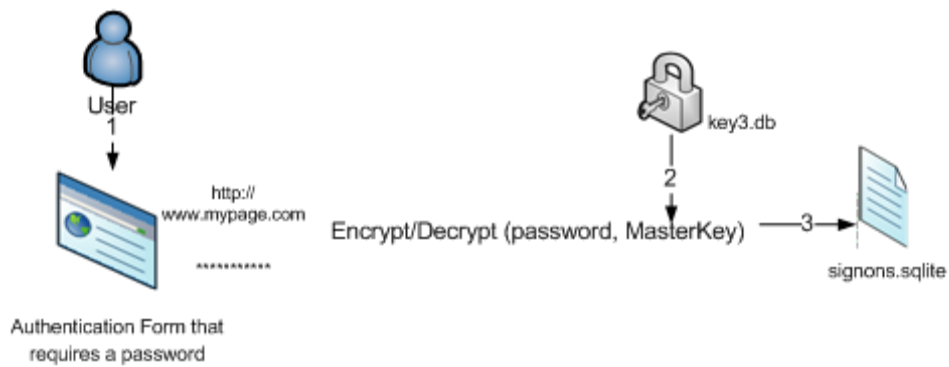

*Figure 3. Firefox 3.6 solution for storing AutoPasswords*

Both of these two files are located on the Firefox profile directory. For a Windows platform the profile directory is:

- Windows 7:
  %APPDATA%\Roaming\Mozilla\Firefox\Profiles\xxxxxxxx.default\
- Vista/XP/2000 platform in
  %APPDATA%\Mozilla\Firefox\Profiles\xxxxxxxx.default\
- Windows 98/Me platform in
  C:\WINDOWS\Application Data\Mozilla\Firefox\Profiles\xxxxxxxx.default\

This is an improved solution than the ones implemented by previous versions of Firefox, [2], [6]. Before using SQLite, [2], Firefox stored passwords in text files named *signons.txt* before Firefox 1.5, *signons2.txt* for version 2 and *signons3.txt* for Firefox version 3.

Using an ASCII editor to open *signons.sqlite* file, you can find the sql create command for the moz_logins table:

*CREATE TABLE moz_logins (id INTEGER PRIMARY KEY,hostname TEXT NOT NULL, httpRealm TEXT,formSubmitURL TEXT, usernameField TEXT NOT NULL, passwordField TEXT NOT NULL, encryptedUsername TEXT NOT NULL, encryptedPassword TEXT NOT NULL, guid TEXT, encType INTEGER).*

A sample record is:
*http://www.doctorat.ase.rohttp://www.doctorat.ase.roTextBoxNumeTextBoxParolaMDIEEPgAAAAAAAAAAAAAAAAAAAEwFAYIKoZIhvcNAwcECAExnKf8hO/BBAgu8/xGl4o27w==MDoEEPgAAAAAAAAAAAAAAAAAAEwFAYIKoZIhvcNAwcECN1e72k2a8qDBBCvfgx7w+yy6BJDcPXEXSph{2a43307a-2a96-45ac-a79d-942de8283c06}*
and it can be interpreted as in table 1:

Table 1. Example of Firefox password manager records.

| Field name | Value | Description |
| --- | --- | --- |
| id | | The record id; in this example its value is 1 in binary |
| hostname | http://www.doctorat.ase.ro | Website name |
| httpRealm | - | Not used |
| formSubmitURL | http://www.doctorat.ase.ro | The URL of the Web form used to authenticate users |
| usernameField | TextBoxNume | The name of the username HTML field |
| passwordField | TextBoxParola | The name of the password HTML field |
| encryptedUsername | MDIEEPgAAAAAAAAAAAAAAAAAAAEwFAYIKoZIhvcNAwcECAExnKf8hO/BBAgu8/xGl4o27w== | Encrypted and Base64 encoded value of the username |
| encryptedPassword | MDoEEPgAAAAAAAAAAAAAAAAAAEwFAYIKoZIhvcNAwcECN1e72k2a8qDBBCvfgx7w+yy6BJDcPXEXSph | Encrypted and Base64 encoded value of the password |
| guid | {2a43307a-2a96-45ac-a79d-942de8283c06} | unique GUID for the record |
| encType | | Binary 1 value representing an encrypted record |

*Table 2. Firefox data management solutions.*

| File name | Data contained |
|---|---|
| Places.sqlite | Last visited URLs in the *moz_places* table<br>Bookmarks in the *moz_bookmarks* table related to the last visited URLs<br>Address bar input history in *moz_inputhistory* table |
| Downloads.sqlite | Downloaded files, their URL and the destination directory in the *moz_downloads* table |
| Formhistory.sqlite | Autocomplete form data in the *moz_formhistory* table; each record contains the form field name and its value |

Firefox saves other sensitive related Internet data, as specified in this paper, but it doesn't encrypt it. It uses *SQLite* files to store clear text representing categories described in table 2. As seen from table 1 and 2, an attacker has no problems on acquiring sensitive data, other than the AutoPasswords, because the information in stored in clear text in known locations. For gaining access to the *signons.sqlite* data, an attacker must get the master password, stored in k*ey3.db*, used to encrypt user passwords. There are different approaches to do this:

▪ Brute force by generating all possible passwords from a given vocabulary; for a secure password this is almost impossible as it will require a very long period of time;
▪ Dictionary by trying common passwords; the attack is based on the user commodity and the belief that his password is something easy to remember and significant
▪ Hybrid by using generated prefixes and suffixes with common dictionary passwords;

and in practice, these solutions has been implemented by different software applications, [6-7], [30]. Based on that, we can conclude that the security of the Firefox password manger lies entirely on the strength of the master password. But, in many situations, without an imposed security policy, the user saves its passwords without setting a master password.

# 4. Internet Explorer security

At the time this research is conducted, the latest version for Internet Explorer is 8, and despite that, a large part of Internet users, 15% [21], still use earlier versions, like 6. For this version of IE the passwords are stored in a secret location in the Registry known as the Protected Storage, [19].

The base key used to encrypt data from the Protected Storage is located under the following key *HKEY_CURRENT_USER\Software\Microsoft\Protected Storage System Provider* and is related to the user Windows account.

From 7 and 8 versions, Internet Explorer stores sensitive data also in Registry, but in a more secure way, [8] [18-19]:

▪ AutoComplete passwords are stored in the Registry under *HKEY_CURRENT_USER\Software\Microsoft\Internet Explorer\IntelliForms\Storage2*.
▪ HTTP Authentication passwords are stored in the Credentials file under *Documents and Settings\Application Data\Microsoft\Credentials*.

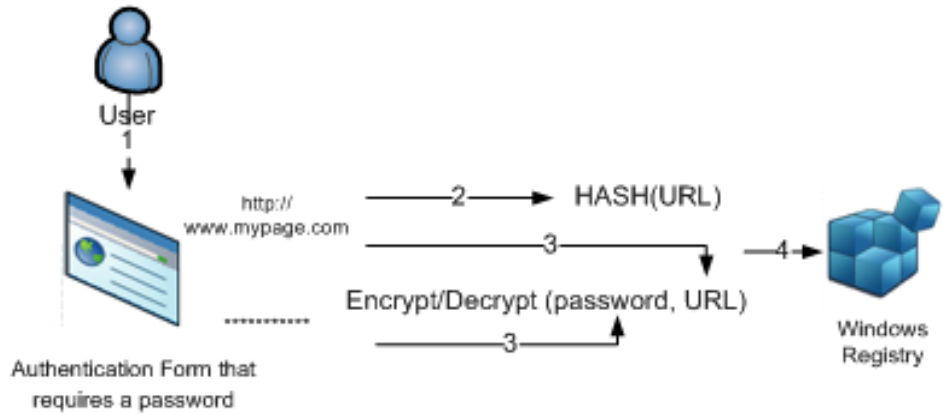

*Figure 4. Internet Explorer solution for storing AutoPasswords.*

As seen in figure 4 the passwords stored in registry are encrypted using a procedure [18-19] that uses relative information for each encryption and not an absolute master key. Figure describes the encryption/decryption steps:

1. The user access the Web page that uses a HTML authentication form;

2. After authentication, if the user chooses to save the password and username, Internet Explorer stores the page's URL hashing it using SHA1 hash function;

3. The URL is used as a key to encrypt the password with strong encryption algorithms provided by the Windows OS through Data Protection Application-Programming Interface (DPAPI);

4. The pair-values, URL hash and encrypted password, are added to a new record in the registry

5. The URL is not needed in clear text for decryption because it is obtained every time the user accesses the Web page.

The encryption of Autoform data is implemented a bit different way [18-19] because instead of the URL the software uses the HTML form field name for the encryption key.

The decryption process is the almost the same as the encryption one. The only difference comes from the fact that every time the user access a Web page that contains an authentication form, the hash of the URL is compared with every record in the registry in order to determine if there are stored passwords.

At the first glance, the process has a high security level because the encryption key is not required to be stored on the local computer. In reality, this is achieved only if the user deletes browsing history from the computer after using Internet Explorer.

From an attacker point of view it is almost impossible to guess the page URL based on its hash. So he will not be able to decrypt the registry records without other information:

▪ the attacker tries to find the user password for a particular Website, so he compares the URL hash with the registry records; if he finds a match, he will be able to decrypt the password;

▪ the attacker tries to break all AutoComplete passwords for that user; the additional needed information is found for most users in the browsing history data because it contains the URL of all visited Web pages, including the ones that require HTML authentication; the attacker will generate the hash for each URL and compares it with the ones in registry; for each match, the attacker will decrypt the password; because most users don't delete browsing history, they are very vulnerable to this kind of attack; also IE password recovery software,[10][18][19] implements this solution;

Other sensitive data as bookmarks, visited URLs, cookies are stored in clear at different locations. For Windows Vista or 7, these text files can be found at:

- Cookies in this folder - *C:\Users\<username>\AppData\Roaming\Microsoft\Windows\Cookies*
- History in this folder - *C:\Users\<username>\AppData\Local\Microsoft\Windows\History*
- Favorites in this folder - *c:\Users\<username>\favorites*

## 5. Google Chrome

For the Google Chrome Web browser, version 7 at the research moment, the passwords are stored in:

- Windows Vista/XP: *[Windows Profile]\AppData\Local\Google\Chrome\User Data\Default\Web Data*
- Windows 7: *[Windows Profile]\AppData\Local\Google\Chrome\User Data\Default*

as records in SQLite database file formats which contains encrypted passwords and other sensitive user data.

Figure 5 describes the process of encrypting and decrypting user passwords, [23], by Google Chrome in 3 steps:

1. The user access the Web page that uses a HTML authentication form;
2. After authentication, if the user chooses to save the password and username, Chrome stores; for encryption, Chrome uses Windows Data Protection application programming interface (DPAPI) functions, [22]; the encryption key is determined by the Windows user logon credentials;
3. The encrypted form of the password and username are stored in *Login Data* file, which has a SQLite format.

*Login Data* file can be easily interpreted because it SQLite structure, described in table 3, but the password is the only field which has its value encrypted.

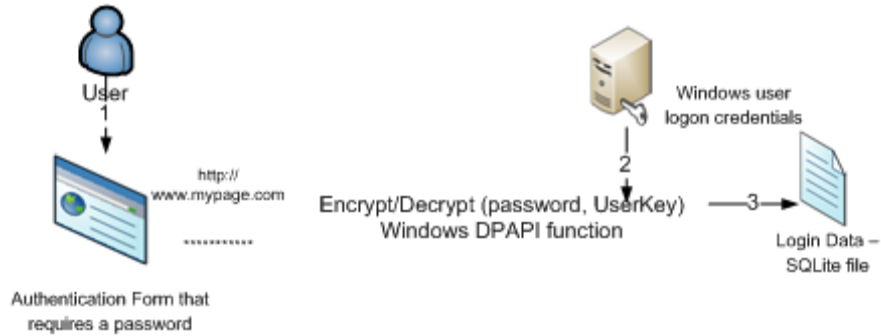

*Figure 5. Google Chrome solution for storing AutoPasswords*

*Table 3. Example of Chrome password manager records.*

| Field name | Value | Description |
|---|---|---|
| rowid | 1 | The record id; in this example its value is 1 in binary |
| origin_url | http://www.doctorat.ase.ro/Log/Default.aspx | The URL of the Web form used to authenticate users |
| action_url | http://www.doctorat.ase.ro/Log/Default.aspx | The URL of the Web form used to authenticate users |
| httpRealm | - | Not used |
| formSubmitURL | http://www.doctorat.ase.ro | The URL of the Web form used to authenticate users |
| username_element | TextBoxNume | The name of the username HTML field |
| username_value | catalin | The value of the username in |

| | | plaintext |
|---|---|---|
| **password_element** | TextBoxParola | The name of the password HTML field |
| **password_value** | BLOB (Size: 230) | Encrypted value of the password |
| **submit_element** | - | - |
| **signon_realm** | http://www.doctorat.ase.ro | Website name |
| **ssl_valid** | 0 | Binary 1 value representing an SSL connection |
| **preferred** | - | - |
| **date-created** | - | Date of the record |

*Table 4. Chrome data management solutions.*

| File name | Data contained |
|---|---|
| **Web Data** | SQLite file that contains Autocomplete form data in the *autofill* table; each record contains the form field name and its value<br>Credit card data for e-payment Web forms in the *credit_cards* table |
| **Bookmarks** | Bookmarks related to the last visited URLs; the file has a simple ASCII content and uses a structure based format |
| **History** | Last visited URLs in the *urls* table<br>Number of visits in the *visits* table<br>Downloaded files, their URL and the destination directory in the *downloads* table<br>Used keyword search terms in *keyword_search_terms* |
| **Top Sites** | Most visited URLs in the *thumbnails* table |

Chrome saves other sensitive related Internet data, but in a clear text format. Table 4 describes the data managed by Google Chrome and how it is stored.

Except from the user password, any other sensitive data is managed in plaintext and can be accessed with ease. To gain access to the Autopasswords an attacker must gain access to the user machine and to the Windows user credentials. These constraints are generated by the CryptProtectData function, [29] implementation, from the Windows DPAPI, [22]. Once the attacker has logged into user Windows account, he has complete access to the passwords.

## 6. Comparative analysis

In an independent security test done by NSS Labs in January 2010, [11], Internet Explorer obtained by far best results. The study also took into consideration results obtained in the 2009 third quarter and made a security improvement assessment based on that.

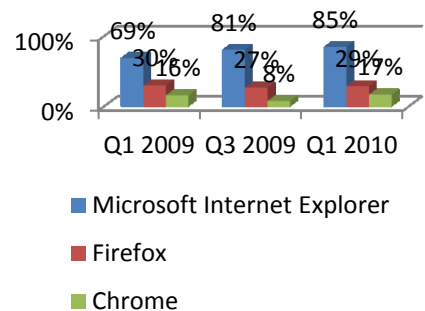

*Figure 6. NSS Labs malware security test – Q1 2010 [11].*

Accordingly to the tests specifications, the browsers security has been evaluated against a large set of socially-engineered malware examples. Accordingly to the figure 6 results, Internet Explorer has obtained by far the best results.

Based on the security analysis of the three Internet browsers and a previous feature comparative analysis in [2], table 5 compares their password manager security options.
Considering the vulnerabilities of each presented solution, table 6 describes some attack types and their success probability for gaining access to the password database.

Table 5. Comparison of password management security options.

| Option | IE 8 | Firefox 3.6 | Chrome 7 |
|---|---|---|---|
| Storage of usernames, passwords, and URLs | x | x | x |
| Password access via JavaScript | x | x | x |
| Password access via software program | | x | x |
| Password Protection not tied to Windows user account | | x | |
| Password protection based on a single master password | | x | x |
| Password Prompt when starting session to use stored web passwords | x | x | x |
| Easily Exportable username/password data | x | x | |
| Encrypted | x | x | X |
| Encoded | x | x | |
| Password Manager option to "Show Passwords" in clear | | x | X |

Table 6. Comparison of attacks types on browser password management data.

| What the attacker does to get the passwords | IE 8 | Firefox 3.6 | Chrome 7 |
|---|---|---|---|
| Running a javascript/booklet on the Web form | x | x | x |
| Installs malware on the target computer using user account | x | x | x |
| Uses XSS on the target Website | x | x | x |
| Get Windows user credentials | | | x |
| Get URL history | x | | |
| Brute force/Dictionary/Hybrid attack in the master password | | x | |
| Social engineering attacks | x | x | x |
| Key logging | x | x | |
| Copies the password management files | | x | x |
| Can use a GUI function to view password database | | x | x |

Each solution has its own advantages and disadvantages but none offers guaranteed security and their strength depends on how are used.

# 7. Conclusion

To protect its sensitive data and to enhance the security of his computer, a user must update and patch its application, to use browser security add-ons and anti-malware/anti-email spam software. Also he must educate himself not to disclose it so easy and to distinguish fake requests.

Another solution is to embrace a Security by Isolation approach, [16] and use The Red, Yellow and Green virtual machines to separate daily browsing from online shopping and from accessing the online banking account.

In order to reduce security risks, users could use additional software, integrated in antivirus or standalone that will erase browsing history and other clear text data on a high frequency basis. In secure working environments, the browsers usability and accessibility related functions are less important than the data security and become sources of security vulnerabilities. An alternative to browsers password managers are the standalone applications or plug-ins, [15].

## Acknowledgements

The research leading to this paper has been supported by The European Social Fund through Sectored Operational Programmer Human Resources Development 2007-2013, project number POSDRU/89/1.5/S/59184, "Performance and excellence in postdoctoral research in Romanian economics science domain".

Parts of this research have been published in the Proceedings of the 3rd International Conference on Security for Information Technology and Communications, SECITC 2010 Conference (printed version).